%% file: mnras.tex
\documentclass[fleqn,usenatbib]{mnras}
\usepackage{fix-cm}

\usepackage{newtxtext,newtxmath}

\usepackage[T1]{fontenc}
\usepackage{algorithm,algpseudocode}
\usepackage{multirow}
\usepackage{tabularx}
\usepackage{fontawesome5}
\DeclareRobustCommand{\VAN}[3]{#2}
\let\VANthebibliography\thebibliography
\def\thebibliography{\DeclareRobustCommand{\VAN}[3]{##3}\VANthebibliography}

\newcommand{\Mpch}{$h^{-1}\,\mbox{Mpc}$}

\newcommand{\MpchInv}{$\mbox{Mpc}/h$}

\usepackage{graphicx}	
\usepackage{amsmath}	
\usepackage{multirow}
\usepackage{colortbl}
\usepackage{ulem}

\title[Generation of 3D dark matter density fields]{Conditional Diffusion-Flow models for generating 3D cosmic density fields: 
applications to $f(R)$ cosmologies\\ 
}

\author[Riveros et al.]{
Julieth Katherine Riveros,$^{1}$
Paola Saavedra,$^{1}$
H\'ector J. Hort\'ua, $^{1,2}$\thanks{E-mail: H.J.HortuaOrjuela@ljmu.ac.uk} \and
Jorge Enrique García-Farieta,$^{3}$
and Ivan Olier$^{2}$
\\
$^{1}$Grupo Signos, Departamento de Matem\'aticas, Universidad El Bosque, Bogot\'a, Colombia \\
$^{2}$ Data Science Research Centre, Liverpool John Moores University, 3 Byrom Street, Liverpool L3
3AF, UK\\
$^{3}$ Departamento de F\'isica, Universidad de C\'ordoba, E-14071, Córdoba, Spain
}

\begin{document}
\label{firstpage}
\pagerange{\pageref{firstpage}--\pageref{lastpage}}
\maketitle

\begin{abstract}
Next-generation galaxy surveys promise unprecedented precision in testing gravity at cosmological scales.  However, realising this potential requires accurately modelling the non-linear cosmic web.  We address this challenge by exploring conditional generative modelling to create 3D dark matter density fields via score-based (diffusion) and flow-based methods. Our results demonstrate the power of diffusion models to accurately reproduce the matter power spectra and bispectra, even for unseen configurations. They also offer a significant speed-up with slightly reduced accuracy, when flow-based reconstructing the probability distribution function, but they struggle with higher-order statistics. To improve conditional generation, we introduce a novel multi-output model to develop feature representations of the cosmological parameters. Our findings offer a powerful tool for exploring deviations from standard gravity, combining high precision with reduced computational cost, thus paving the way for more comprehensive and efficient cosmological analyses~\href{https://github.com/JavierOrjuela/generative-models-f_R_2025}{\faGithub}.
\end{abstract}




\section{Introduction}

The advent of precision cosmology marks a new era in our understanding of the Universe driven by a variety of upcoming missions. Among the key ongoing and forthcoming efforts are the Dark Energy Spectroscopic Instrument (DESI) \citep{2016arXiv161100036D}, the Euclid space mission \citep{2011arXiv1110.3193L,2018LRR....21....2A}, the Legacy Survey of Space and Time (LSST) \citep{2020arXiv200907653V}, the Wide-Field Infrared Survey Telescope (WFIRST) \citep{2019arXiv190205569A}, and the Square Kilometre Array (SKA) \citep{2020PASA...37....7S}. These experiments aim to provide unprecedented measurements that will constrain cosmological parameters with high precision, with N-body simulations being a crucial component of these efforts. N-body simulations are, in fact, essential for accurately modelling the large-scale structure of the universe, understanding the evolution of cosmic fields, and interpreting the data from these surveys. N-body simulations play a fundamental role in understanding the physics behind galaxy survey data, as they enable the exploration of cosmic structures across a range of scales. Although the non-linear regime of structure formation can, in principle, be approximated by perturbative-based methods \citep[for different approaches, see e.g.][]{2019PhRvD..99f3530O,2014JCAP...01..010B,2019JCAP...11..027K,2023PDU....4001193C,2002PhR...367....1B,2006PhRvD..73f3519C,2009PhRvD..80d3531C,2009PhR...475....1M,2012JCAP...07..051B,2012JHEP...09..082C}, there is currently no single, universally adopted framework. Simulations not only offer insights into the small-scale behaviour of galaxy clustering but also provide a reliable means of investigating the clustering in cosmologies beyond the $\Lambda$-Cold Dark Matter ($\Lambda$CDM) model. As a result, simulations have become increasingly crucial in exploring modified gravity models, with $f(R)$ models highlighting a minimal but significant modification of  Einstein's general relativity (GR). Since deviations from GR are likely to manifest in summary statistics, accurately predicting these statistics is crucial for connecting to theoretical models of structure formation. Despite their advantages, simulations can be computationally expensive, depending on their complexity, such as the richness of physical phenomena included and the resolution of mass and scale. This has led to the rise of emulators based on deep learning algorithms that are designed to quickly and accurately predict cosmological observables. 
In particular, generative models have emerged as a promising tool in cosmology, especially for enhancing and accelerating the analysis of simulations. Recent advancements in generative models offer a powerful approach to efficiently approximate and generate maps where their summary statistics closely mirror the ones obtained from simulated data. By learning the underlying distribution of the complex, high-dimensional cosmic structures, generative models can potentially provide faster and more scalable solutions while maintaining accuracy, opening new avenues for both theoretical and observational cosmology.
Generative models have been widely used in astronomy such as autoencoders~ \citep{2024arXiv240302171U, 2022MNRAS.513..333R, 2021MNRAS.500..531A, 2023ApJ...952..145J,2025MNRAS.537..448S}, normalizing flows~\citep{2022ApJ...937...83H, 2024ApJ...976...76K, 2025MNRAS.536..190M, 2021arXiv210512024R, 2024A&A...684A.100G}, generative adversarial networks (GANs)~\citep{2025MNRAS.536.3138B, 2025MNRAS.536.1408G, 2024arXiv240210997A, 2023arXiv230704976D, 2022arXiv221105000A, 2022JCAP...12..013Y, 2020arXiv200408139P, 2021MNRAS.506.3049T, 2018ComAC...5....4R, 2019ComAC...6....5P, 2025MNRAS.536.1408G, 2021arXiv211106393S, 2019ComAC...6....1M, 2024arXiv240809051Z, 2020MNRAS.495.4227K, 2021A&A...651A..46U}, where the authors have demonstrated them to be powerful tools for simulating samples from complex probability distributions \citep[see][for a comprehensive review]{GM2020100285,010814-020120}. Recently, the potential of diffusion models has gained increasing attention, with several studies exploring their application in cosmology for emulating satellite galaxy and subhalo populations~\citep{2024arXiv240902980N}, field emulation and parameter inference~\citep{2023arXiv231207534M}, emulators~\citep{2024PhRvD.109l3536R,2023arXiv231100833H} and image generation~\citep{mudur2022denoisingdiffusionprobabilisticmodels,mudur2023cosmologicalfieldemulationparameter,zhao2023diffusionmodelconditionallygenerate} among others. Diffusion models have gained significant attention due to their effectiveness in generating high-quality samples \citep[for a review, see][]{10081412,10419041}. These models define a forward diffusion (noising) process that gradually transforms samples from the target distribution into samples from a standard normal distribution. The reverse diffusion process, which is learned during training, is equivalent to learning the data score, which is why diffusion models are also referred to as score-based generative models. This framework has demonstrated remarkable success in photorealistic image generation, as exemplified by Stable Diffusion, and addresses some of the key limitations of GANs, such as mode collapse—where the model fails to capture all modes of the distribution~\citep[see e.g.][]{ho2020denoising}. 
 In this work, we apply diffusion models to generate cold-dark-matter 3D-density fields of modified $f(R)$ gravity conditioned on cosmological parameters. By leveraging conditional diffusion models, we demonstrate their ability to emulate fast and accurate full 3D density fields while maintaining consistency with the summary statistics, all at a low computational cost, with an accuracy similar to state-of-the-art N-body simulations of modified gravity models.
 
The outline of this paper is as follows. In Section 2, we describe three diffusion methods employed in this research and illustrate the algorithms used for training and deploying the models. Section 3 introduces the modified gravity simulations used for training and evaluating the trained-models, and Section 4 we detail our methodology including the neural network, the use of representation learning for including the cosmological parameters as conditioned on the generative models, and the n-point statistics for evaluating the performance of the models. In Section 5, we present the results of the predicted observables for the different methods implemented and assess their performance. Finally, in Sections 6 and 7, we discuss the results and provide the main conclusions of this research.


\section{Preliminaries}
This section outlines the basis for conditional diffusion models as emulators for N-body simulations. Additionally, we introduce several strategies developed to implement diffusion model flavours.
\begin{figure}
    \centering
    \includegraphics[width=\linewidth]{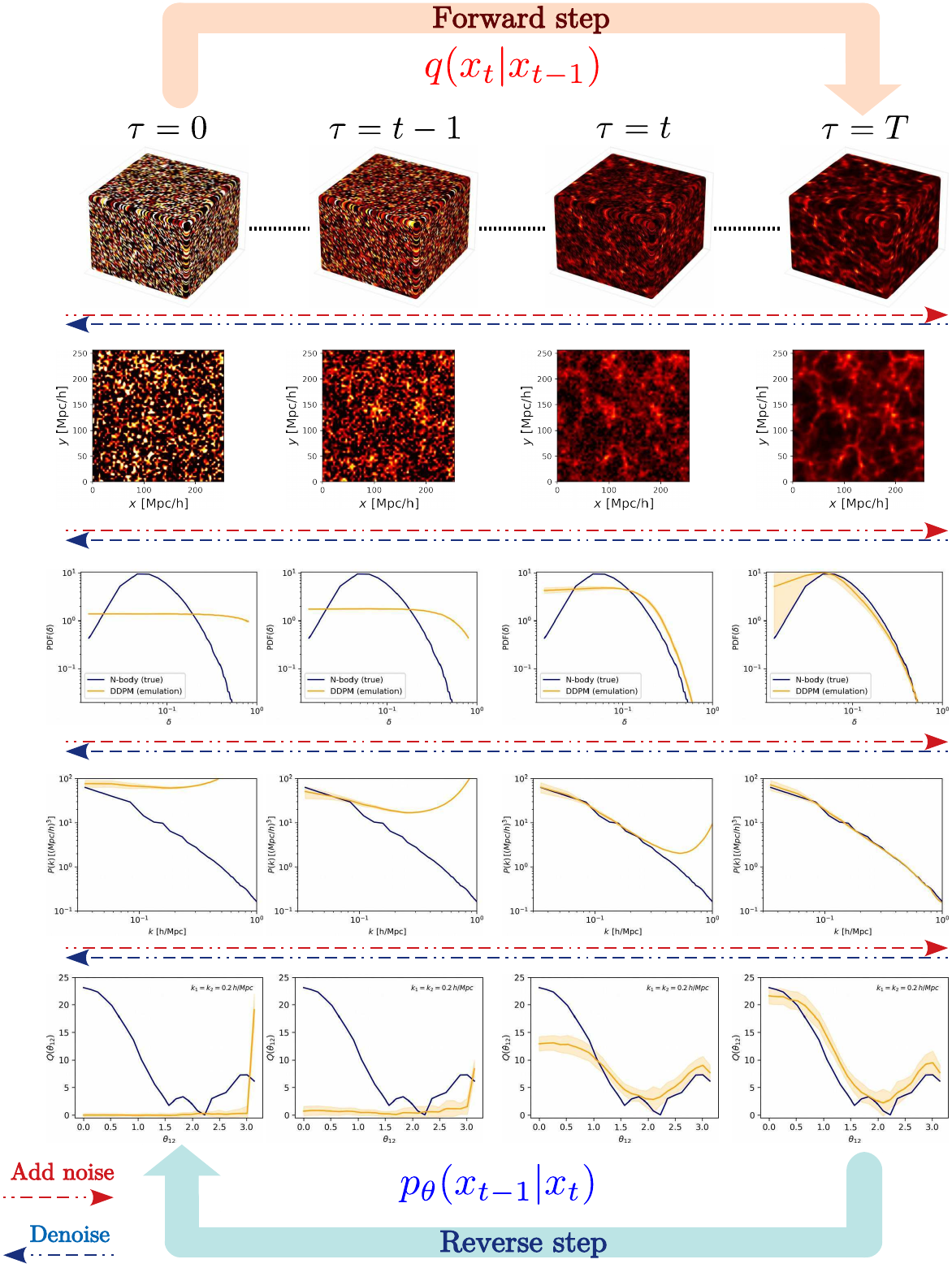}
    \caption{The diagram illustrates the forward and reverse processes in DDPM. The top panel shows the generative 3D density fields created at each time step $\tau$, with their corresponding 2D projections displayed below. The centre panel presents the power spectra (PS), followed by the evolution of one bispectrum configuration. The bottom panel depicts the probability distribution function (PDF) of the voxels (see Sec.~\ref{secmoments} for definitions of these statistical moments). In the plots, the blue curves represent the true statistical moments, while the orange curves correspond to those computed from the generated simulation at each time step $\tau$.}
    \label{fig:diff}
\end{figure}
\subsection{\textbf{Denoising Diffusion Models}}
Diffusion Probabilistic Models (DPMs) have rapidly gained prominence as a highly promising generative technique in recent years. Functioning as latent variable models for sequence modelling, DPMs utilise a latent space with the same dimensionality as the input data. In contrast to Generative Adversarial Networks (GANs)~\citep{goodfellow2014generative}, which are not probabilistic models, DPMs offer significant advantages, including excellent parallelisation capabilities and avoidance of adversarial training~\citep{ho2020denoising}. This eliminates the well-known challenges of debugging and convergence difficulties frequently encountered with GAN training.

\subsection{\textbf{Denoising diffusion probabilistic model (DDPM)}}
Generative models aim to learn and approximate complex, high-dimensional data distributions~\citep{lamb2021briefintroductiongenerativemodels}. Among these, probabilistic diffusion models have recently emerged as a powerful technique, distinguished by their capacity to transform unstructured noise into highly detailed and structured outputs that closely resemble the training data distribution~\citep{ho2020denoising}. This is achieved through a two-stage process: a \textit{forward diffusion process} that progressively adds noise to the data, gradually corrupting its structure, and a \textit{reverse diffusion process} that learns to reverse this corruption, iteratively building coherent structures from the noise. This bidirectional approach, involving both forward and reverse diffusion, has proven particularly effective for challenging astrophysical scenarios, most notably in generating high-resolution fields~\citep{schanz2023stochasticsuperresolutioncosmologicalsimulations},reconstruction~\citep{sabti2024generativemodelingapproachreconstructing}, and as emulators~\citep{mudur2024diffusionhmcparameterinferencediffusion,zhao2023diffusionmodelconditionallygenerate,mudur2022denoisingdiffusionprobabilisticmodels},  setting new benchmarks for generative modelling performance. The learning aspect of these models involves mastering the reversal of a complex noising process, where progressively more noise is actively added to an initial image \(x_0 \sim q(x_0)\)
~\citep{schanz2023stochasticsuperresolutioncosmologicalsimulations}. This noise sequence is executed through a Markov chain of \(T\) steps and systematically introduces Gaussian noise at each stage, generating a sequence of noise samples \( x_1,..., x_t\). According to \cite{ho2020denoising}, during the forward diffusion step, noise is introduced to a sample \( x_t \) from the preceding one \( x_{t-1} \) and step sizes are regulated by a variance $\{ \beta_t \in (0, 1) \}_{t=1}^T$ as
\begin{equation}\label{equation1}
    q(\mathbf{x}_t | \mathbf{x}_{t-1}) = \mathcal{N}\left(\mathbf{x}_t ; \sqrt{1 - \beta_t} \mathbf{x}_{t-1}, \beta_t \mathbf{I}\right).
\end{equation}
where $\beta_{1}<\beta_{2}<\cdot\cdot\cdot<\beta_{t}$ regulates the reduction in noise between steps. In Eq.~(\ref{equation1}) the process assumes that \( x_t \) is conditionally Gaussian with a mean \(\sqrt{1 - \beta_t} x_{t-1}\) and variance \(\beta_t \mathbf{I}\), where \(\mathbf{I}\) is the identity matrix. The mean term controls how much of \( x_{t-1} \) contributes to \( x_t \), while the variance introduces isotropic Gaussian noise with a magnitude determined by \(\beta_{t}\)~\citep{weng2021diffusion}. Therefore, by repeatedly applying the forward diffusion process, the image at a specific time \( t \), denoted \( x_t \), can be expressed as a function of the original field \( x_0 \). From conditional probabilities, the following joint probability is calculated as
\begin{equation}\label{equation2}
    q(x_{1:T} \mid x_0) = \prod_{t=1}^{T} q(x_t \mid x_{t-1}).
\end{equation}
The term \(q(x_{1:T} \mid x_0)\)  represents the overall probability of observing the sequence \( x_1 \) to \( x_T \). Each factor \(q(x_t \mid x_{t-1})\) denotes the probability of transitioning to state \( x_t \) from the previous state \( x_{t-1} \), capturing the Markov property, where the next state is determined solely by the current state~\citep{weng2021diffusion}. A notable feature of this process is that sampling at any arbitrary time step can be achieved in closed form by leveraging the reparameterization trick ~\citep{ho2020denoising}. This property allows for direct access to any sample \(x_t\) eliminating the need to sequentially compute all \( t-1\)  previous noisy image
\begin{equation}\label{imagesbef}
x_t = \sqrt{\bar{\alpha}_t}x_0+\sqrt{1-\bar{\alpha}_t}\epsilon,
\end{equation}
where $\alpha_t=1-\beta_t$, with $\bar{\alpha}_t=\Pi_i^t\alpha_i$, and $\epsilon\sim\mathcal{N}(\mathbf{0},\mathbf{I})$. The noise added at each step is systematically removed during the reverse diffusion phase~\citep{ho2020denoising}. As a result, the process begins with a distribution that contains only noise (the final state of the forward process). Consequently, the noise is removed from the samples step by step, moving in the reverse direction.  As stated in \cite{ho2020denoising}, the inverse diffusion process considers \( x_0 \), and the events are connected through the conditional probability distribution
\begin{equation}\label{equation3}
p_\theta(\mathbf{x}_{0:T}) = p(\mathbf{x}_T) \prod_{t=1}^{T} p_\theta(\mathbf{x}_{t-1} \mid \mathbf{x}_t).
\end{equation}
being the reverse process equals to
\begin{equation}\label{equation4}
p_\theta(\mathbf{x}_{t-1} \mid \mathbf{x}_t) = \mathcal{N}(\mathbf{x}_{t-1}; \mu_\theta(\mathbf{x}_t, t), \Sigma_\theta(\mathbf{x}_t, t)).
\end{equation}
Here, a neural network with parameters $\theta$ is used to  compute  Eq~(\ref{equation3}).  This express the joint probability distribution $p_\theta(\mathbf{x}_{0:T})$ over a sequence of variables \( x_0, x_1,..., x_T\) parameterized by \(\theta\) ~\citep{weng2021diffusion}. We can divide this joint probability into two parts: the marginal probability of the final state \( p(x_T) \), and the product of conditional probabilities \( p_{\theta}(x_{t-1} \mid x_t)\)  over all timesteps \( t \) from 1 to \( T \)~\citep{weng2021diffusion}. This structure reflects a reverse process, where each state \( x_{t-1}\) depends on the subsequent step \( x_{t} \). The variance is usually selected as $\Sigma_\theta(\mathbf{x}_t, t)=\beta_t\mathbf{I}$ as~\citet{ho2020denoising} reported to be the best performance in their results, while the mean $\mu_\theta(\mathbf{x}_t, t)$ is given by
\begin{equation}\label{x_reserve}
  \mu_\theta(\mathbf{x}_t, t)= \frac{1}{\sqrt{\alpha_t}}\big(x_t-\frac{\beta_t}{\sqrt{1-\bar{\alpha}_t}}\epsilon_\theta(\mathbf{x}_t, t) \big),
\end{equation}
where $\epsilon_\theta(\mathbf{x}_t, t)$ is the neural network outcome of the noise $\epsilon$ present in the sample $x_t$.   The loss function used for this optimisation is given by the expectation value~\citep{ho2020denoising}
\begin{equation}\label{equation7}
\mathcal{L}_{t} =\mathcal{E}_{t\sim[1,T],x_0,\epsilon}\bigg[\big\| \boldsymbol{\epsilon} - \boldsymbol{\epsilon}_\theta \big( \mathbf{x}_t, t \big) \big\|^2\bigg],
\end{equation}
being $\boldsymbol{\epsilon}_\theta \big( \mathbf{x}_t, t \big)$ the neural network prediction of the noise $\epsilon$ present in the sample $x_t$, and $t\sim[1,T]$  the time step drawn from a uniform distribution.  By minimising this loss, the model learns to predict and remove the noise at each step, enabling it to reverse the diffusion process during inference and generate realistic data from random noise. 
The training algorithm is listed in Algorithm~\ref{algo1}. Once training is completed, we expect to generate $x_0\sim p(x_0)$ image from noise.  In fact, the model learns to approximate the probability distribution of the training set. Hence, we can sample from this
distribution and be able to generate new samples that obey the same features as the training dataset. This can be done by sampling $T$ times Eq.~\ref{equation4} crossing the Markov chain until $t=0$ as
\begin{equation}\label{inferen_x_0}
  x_{t-1}= \frac{1}{\sqrt{\alpha_t}}\big(x_t-\frac{\beta_t}{\sqrt{1-\bar{\alpha}_t}}\epsilon_\theta(\mathbf{x}_t, t) \big) + \sqrt{\beta_t}z,
\end{equation}
with $z\sim\mathcal{N}(\mathbf{0},\mathbf{I})$. Here, the first term is the mean estimate provided by the neural network Eq.~\ref{x_reserve} perturbed by the presence of a  Gaussian noise $\beta_t$ akin to a Langevin sampling step~\citep{Welling2011BayesianLV}. The inference algorithm is listed in Algorithm~\ref{algo2}.
\begin{figure}
\begin{algorithm}[H]
  \caption{DDPM Training}\label{algo1}
  \begin{algorithmic}[1]
     \State Randomly select a simulation $x_0$ and its cosmological parameters $y$ from  the training dataset distribution $q(x_0)$.
     \State Drawn sample from Uniform distribution $\gamma\sim U(0,1)$
     \If{$\gamma<0.1$}
         \State Discard conditioning from the dataset $p(x_0,y=\emptyset)$
     \Else \State Keep conditioning from the dataset $p(x_0,y)$
     \EndIf
     \State Randomly select a time step $t$ in the Markov chain from the uniform distribution $\{1,..., T\}$.
     \State Drawn sample from a Gaussian noise $\mathbf{\epsilon} \sim \mathcal{N} (\mathbf{0}, \mathbf{I})$.
     \State Compute the sample $x_t$ in the $t$-th step of the Markov chain as Eq.~\ref{imagesbef}.
     \State Make a gradient descent step with $\nabla_{\theta}\mathcal{L}_{t}$ defined in Eq.\ref{equation7}.
     \State Repeat steps 1-5 until converged.
  \end{algorithmic}
\end{algorithm}
\end{figure}

\begin{figure}
\begin{algorithm}[H]
  \caption{DDPM Sampling}\label{algo2}
  \begin{algorithmic}[1]
    \State Drawn sample from a Gaussian noise $\mathbf{\epsilon} \sim \mathcal{N} (\mathbf{0}, \mathbf{I})$.
    \State Choose $\omega=[0,1]$: guidance strength.
    \State Loop through the backward Markov chain: 
    \For{$t=T, ..., 1$}
        \If{$t > 1$}
            \State $\epsilon \sim \mathcal{N}(\mathbf{0}, \mathbf{I})$
        \Else \State $\epsilon = 0$ \Comment{No additional noise in the last step}
        \EndIf
        \State Compute $x_{t-1}$ Eq.~\ref{inferen_x_0} where the score estimation is given by Eq.~\ref{free_class}
    \EndFor
    \State Return $x_0$.
  \end{algorithmic}
\end{algorithm}
\end{figure}

\begin{figure}
\begin{algorithm}[H]
  \caption{DDIM Sampling}\label{algo3}
  \begin{algorithmic}[1]
    \State Create  a time subset $\{t_1,..,t_s\}\in \{t_1,..,t_T\} $ with $s\ll T$
    \State Choose $\omega=[0,1]$: guidance strength.
    \State Drawn sample from a Gaussian noise $\mathbf{\epsilon} \sim \mathcal{N} (\mathbf{0}, \mathbf{I})$.
    \State Loop through a subset of timesteps: 
    \For{$t=s, ..., 1$}
        \If{$s > 1$}
            \State $\epsilon \sim \mathcal{N}(\mathbf{0}, \mathbf{I})$
        \Else \State $\epsilon = 0$ \Comment{No additional noise in the last step}
        \EndIf
        \State Compute $x_{t-1}$ Eq.~\ref{inferenceddim} with $\sigma_t=0$ and where the score estimation is given by Eq.~\ref{free_class}
    \EndFor
    \State Return $x_0$.
  \end{algorithmic}
\end{algorithm}
\end{figure}
\subsection{\textbf{Denoising diffusion implicit models (DDIM)}}\label{ddimsec}
DDIMs are implicit probabilistic models associated with DDPMs, since they are trained using the same loss function~\citep{song2022denoising}. DDIMs present an optimised version of DDPM and offer a more efficient and faster solution to the image generation problem. Although it uses the same training objective as DDPM, DDIM introduces non-Markov processes instead of strictly following the Markov approach. This allows DDIM to balance between the quality of the generated samples and processing time. Furthermore, it can create high-quality images faster than DDPM and it performs direct interpolations in latent space and reconstructs observations with minimal error, providing greater flexibility in the generation process~\citep{song2021denoising}.  According to \cite{song2021denoising}, the non-Markov inference process is employed in this case, which leads to the same function applied in the DDPM model mentioned above in equation (\ref{equation1}). Therefore, the DDIM model generalises the  DDPM model and, in turn, allows modifications to the design of the inverse Markov chains. The expression for the
non-Markovian conditional probability distribution  \( p_(x_{t-1} \mid x_t, x_0)\) is
\begin{align}\label{ddimeq1}
p(x_{t-1} \mid x_t, x_0) &= \mathcal{N}( x_{t-1}; \sqrt{\bar{\alpha}_{t-1}}x_0 \nonumber \\
&+ \sqrt{1-\bar{\alpha}_{t-1}-\sigma_t^2}\frac{x_t-\sqrt{\bar{\alpha}_t} x_0}{\sqrt{1-\bar{\alpha}_t}},\sigma^2\mathbf{I}).
\end{align}
According to \cite{song2021denoising}, the processes for the implicit DDIM diffusion models are defined in two phases. In the first phase, the forward diffusion process defines \( x_0 \) and transforms it into \( x_t \). Initially, the inference distribution, the non-Markovian forward process is as follows
\begin{equation}\label{equation8}
p(x_{1:T} \mid x_0) = p(x_T \mid x_0) \prod_{t=2}^{T}  
p(x_{t-1} \mid x_t, x_0),
\end{equation}
where $p( x_{1:T} \mid x_0 )$ corresponds to the conditional probability of observing the sequence of variables \(x_{1:T}\) evolves from an initial state \( x_0 \). It shows that the likelihood of a sequence of observations given the initial conditions can be decomposed in terms of a chain of probabilistic dependencies over time~\citep{zhang2023coherent}. Eq.~\ref{ddimeq1} can be expressed as
\begin{equation}\label{inferenceddim}
    x_{t-1}=  \sqrt{\bar{\alpha}_{t-1}}\frac{x_t-\sqrt{1-\bar{\alpha}_t}\epsilon_\theta(\mathbf{x}_t, t) }{\sqrt{\bar{\alpha}_t}}+\sqrt{1-\bar{\alpha}_{t-1}-\sigma_t^2}\epsilon_\theta(\mathbf{x}_t, t)+\sigma_tz,
\end{equation}
with $z\sim\mathcal{N}(\mathbf{0},\mathbf{I})$, $\epsilon_\theta(\mathbf{x}_t, t)$ is the predicted  neural network noise $\epsilon_t$ at time $t$, and $\sigma_t$ is a parameter learning whose variation determines the difference in the posterior distribution. When $\sigma_t=0$, there is not random sampling and the sample is  generated into a deterministic scenario. This is the core of DDIM model. Besides, since it does not need to satisfy the Markov process, a subset $\{t_1,..,t_s\}\in \{t_1,..,t_T\} $ with $s\ll T$ can
be created from the original $T$  diffusion time-steps for sampling inference,
where $s$ is the number of steps in the new diffusion subset. The inference DDIM algorithm is listed in Algorithm~\ref{algo3}. DDPM and DDIM primarily differ in their approach to sampling. While DDPM relies on a Markov process and requires many diffusion steps to achieve high-quality results, it tends to be computationally expensive~\citep{weng2021diffusion}. The DDIM model offers several improvements over DDPM. It can produce higher-quality samples in fewer steps, enhancing efficiency~\citep{weng2021diffusion}. Moreover, DDIM maintains a consistency property due to its deterministic generative process, ensuring that samples conditioned on the same latent variable share similar high-level features ~\citep{weng2021diffusion}. This consistency also enables DDIM to perform meaningful semantic interpolation within the latent space, resulting in smoother and more interpretable transitions between samples~\citep{weng2021diffusion}.
\subsection{Conditioned Generation: Classifier-Free diffusion guidance}
While training generative models on the simulation, it is important to generate samples conditioned on the cosmological parameters. To explicit incorporate parameter information into the diffusion process, we employ the Classifier-Free Guidance (CFG) in our methodology~\citep{ho2022classifierfreediffusionguidance}.
This technique assumes an unconditional denoising diffusion model $p(x)$ parameterized through an estimator $\epsilon_\theta(x_t,t)=\epsilon_\theta(x_t,t,y=\emptyset)$\footnote{Where for the unconditional model we input a null value $\emptyset$ for the class identifier
y when predicting the score.} and a conditional model $p_\theta(x|y)$ parameterized through $\epsilon_\theta(x_t,t,y)$. Both models are trained with the same neural network. In fact, the conditional diffusion model is trained on data $(x,y)$, where the conditioning cosmological parameters $y$ are randomly discarded by $\gamma<0.1$ (being $\gamma$ a sample drawn from an uniform distribution [0,1]) such that the model knows how to generate images unconditionally as well. Therefore, the score estimator can be written as~\citep{ho2022classifierfreediffusionguidance}
\begin{equation}\label{free_class}
    \bar{\epsilon}_\theta(x_t,t,y)=\epsilon_\theta(x_t,t
    ,y)+\omega(\epsilon_\theta(x_t,t)-\epsilon_\theta(x_t,t,y)),
\end{equation}
where $\omega$ is a parameter that controls the strength of the classifier guidance. In our experiments, we found that $\omega=[0,0.5]$ provides a suitable range of values to obtain good results. The authors ~\citep{ho2022classifierfreediffusionguidance} conclude in their studies that the diffusion model needs to have a part dedicated to the unconditional generation task in order to produce classifier-free guided scores effective for sample quality. 
\subsection{Stochastic Interpolants (SI)}
\begin{figure}
    \centering
    \includegraphics[width=\linewidth]{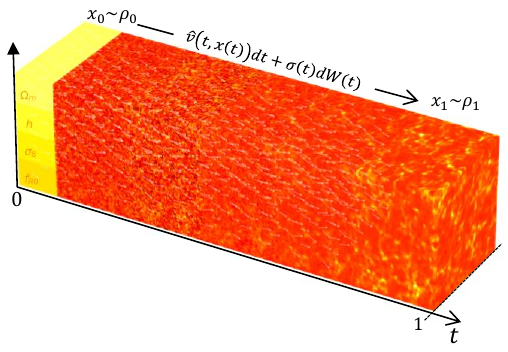}
    \caption{Illustration  of the generative model based on stochastic interpolants, which connects two densities $\rho_0$  and $\rho_1$ that represent the feature representation of the cosmological parameters and the 3D density field respectively. The time-dependent probability density $\rho(t)$ that bridges  $\rho_0$ and $\rho_1$  is found through the the forward stochastic differential equation solutions Eq.~\ref{equation17} where a drift function is computed by the UNet. Here, the vector field represented with white arrows describes the drift function. }
    \label{fig:boxsi}
\end{figure}
Although diffusion-based methods have achieved impressive results in areas such as image generation, there is ongoing research into methods that provide exact transport between arbitrary (not just Gaussian) probability densities within a finite time frame. Initially, score-based diffusion models achieved the best results using Stochastic Differential Equations (SDEs)~\citep{albergo2023stochasticinterpolantsunifyingframework}. However, recent research has shown that methods based on ordinary differential equations (ODE) can achieve comparable or even superior performance if the scoring function is learned effectively. ODE-based methods offer significant advantages, including the availability of an exact and computationally tractable likelihood formula and the straightforward application of established adaptive integration techniques for sampling~\citep{albergo2023buildingnormalizingflowsstochastic}. A recent generative model based on stochastic dynamics propose the use of stochastic interpolants (SI) $x_t$ that connect a base density $\rho_0$ to the target $\rho_1$, but allow for bases that are more general than a Gaussian density. The dynamics can be described as~\citep{albergo2023buildingnormalizingflowsstochastic}
\begin{equation}\label{interpolands}
    x(t)= \alpha(t) x_0 +\beta(t)x_1 +\sigma(t)W(t), \quad t\in[0,1],
\end{equation}
 where by construction, it satisfies $x(t=0)=x_0 \sim \rho_0$, and $x(t=1) = x_1 \sim \rho_1$. This approach therefore exactly bridge between samples from $\rho_0$ at $t = 0$, and from $\rho_1$ at $t = 1$.  For a large class of densities, $\rho_0$ and $\rho_1$ supported on $\mathbb{R}^d$, these distributions are absolutely
continuous with respect to the Lebesgue measure and $\rho(t)$ satisfies  a family of forward and backward Fokker-Planck equations~\citep{albergo2023stochasticinterpolantsunifyingframework}. Therefore, Eq.~\ref{interpolands} can be realized by many different processes such as ODEs and SDEs,  and whose densities at time $t$ are given
by $\rho(t)$~\citep{albergo2023buildingnormalizingflowsstochastic}.
Following the work in~\citep{chen2024probabilisticforecastingstochasticinterpolants,sabti2024generativemodelingapproachreconstructing} let us write the functions under Eq.~\ref{interpolands} as \(\alpha(t) = \sigma(t) = 1 - t\), \(\beta(t) = t^2\), and \(W=\sqrt{t}z\) with $z\sim \mathcal{N}(0,I)$   a Wiener process.  The authors in~\citep{chen2024probabilisticforecastingstochasticinterpolants,albergo2023stochasticinterpolantsunifyingframework} also demonstrate that the velocity field associated with the interpolant, Eq.~(\ref{interpolands}) takes the form 
\begin{equation}\label{equation15}
v(t, x_{0}, x_{1}) = \dot{\alpha}(t)x_{0} + \dot{\beta}(t)x_{1} + \dot{\sigma}(t)W(t),
\end{equation}
where the dot in the variables represents differentiation with respect to time \(t\). The velocity field can be approximately computed with a neural network \(\hat{v}(t, x(t))\) by minimizing the  loss function
\begin{equation}\label{equation16}
\mathcal{L}[\hat{v}] = \int_{0}^{1} dt \, \mathcal{E}_{x_0,x_1\sim \rho_0,\rho_1} \left[ \left( \hat{v}(t, x(t)) - v(t, x_{0}, x_{1}) \right)^2 \right].
\end{equation}
Once trained, the velocity field will function as the drift term within the stochastic differential equation~\citep{sabti2024generativemodelingapproachreconstructing}
\begin{equation}\label{equation17}
\mathrm{d}x(t) = \hat{v}(t, x(t)) \mathrm{d}t + \sigma(t) \mathrm{d}W(t),
\end{equation}
whose solutions are such that $x(t = 1)\sim\rho(x_1|x_0)$ and $W$ accounts for another Wiener process. This equation expresses the evolution of $x(t)$ in terms of two components, the first term describes the deterministic part of the dynamics, while the second term accounts for the stochastic component of the process. To suit this approach to our work, $x_0$ represents a latent representation of the cosmological parameters generated by a neural network while $x_1$ describes the 3D simulation. 
Once the model is trained, the velocity field is substituted by the UNet in Eq.~\ref{equation17}, and the initial volume ($x(t=0)$) is given by the feature representation for the cosmological parameters. The time interval $t\in[0, 1]$ is divided in 200 steps along which we solve Eq.~\ref{equation17}. The Stochastic differential equation was solved using the Euler second-order method. At the end, the emulator should generate distributions of volumes $x(t=1)$ that resemble the characteristics of the density field conditioned on the cosmological parameters.

\begin{figure}
\begin{algorithm}[H]
  \caption{SDEs Training}\label{algo4}
  \begin{algorithmic}[1]
     \State \textbf{Input:} Randomly select a simulation $x_0$ and its labels from  the training dataset distribution $q(x_0)$.
     \State Randomly select a time step $t$ from the uniform distribution $\{1,..., T\}$.
     \State Drawn sample from a Gaussian noise $z \sim \mathcal{N} (\mathbf{0}, \mathbf{I})$ and build the Wiener process. 
     \State Compute $x(t)$ and the velocity field defined in Eqs.~(\ref{interpolands})-(\ref{equation15}) respectively.
     \State Make a gradient descent step with $\nabla_{\theta}\mathcal{L}[\hat{v}]$ defined in Eq.\ref{equation16} and compute it via  Monte Carlo sampling.
     \State Repeat steps 1-5 until converged.
  \end{algorithmic}
\end{algorithm}
\end{figure}

\begin{figure}
\begin{algorithm}[H]
\caption{SDEs Sampling}\label{algo5}
\begin{algorithmic}[1]
\State Randomly select a simulation $x_0$ and its labels from the training distribution,  the trained model $\hat{v}(t_n,x_n)$; and define a  grid $t_0=0<t_1 \cdots< t_{T} =300 $.
\State Set $\Delta t_n = t_{n+1}-t_n$, $n=0:T-1$.
\State Create a 3D image representation of the labels
\State Drawn sample from a Gaussian noise $z_n \sim \mathcal{N} (\mathbf{0}, \mathbf{I})$ and build the Wiener process. 
\State Set $x_1 = x_{0} +  \hat v(t_0,x_{0}) \Delta t_0 + \sigma(t_0) \sqrt{\Delta t_0} z_0$. 
\For{$n = 1:T-1$}
\State Compute $\hat v_{t_n}( x_n, t_n)$ from Eq.~(\ref{equation17}).
\State Set $x_{n+1} = x_{n} +  \hat v(x_n, t_n) \Delta t_n + \sigma(t_n)\sqrt{\Delta t_n} z_n$.
\EndFor
\State \textbf{Return}: $x_T$.
\end{algorithmic}
\end{algorithm}
\end{figure}

\section{Dataset: Modified gravity simulations}
\begin{figure}
    \centering
    \includegraphics[width=\linewidth]{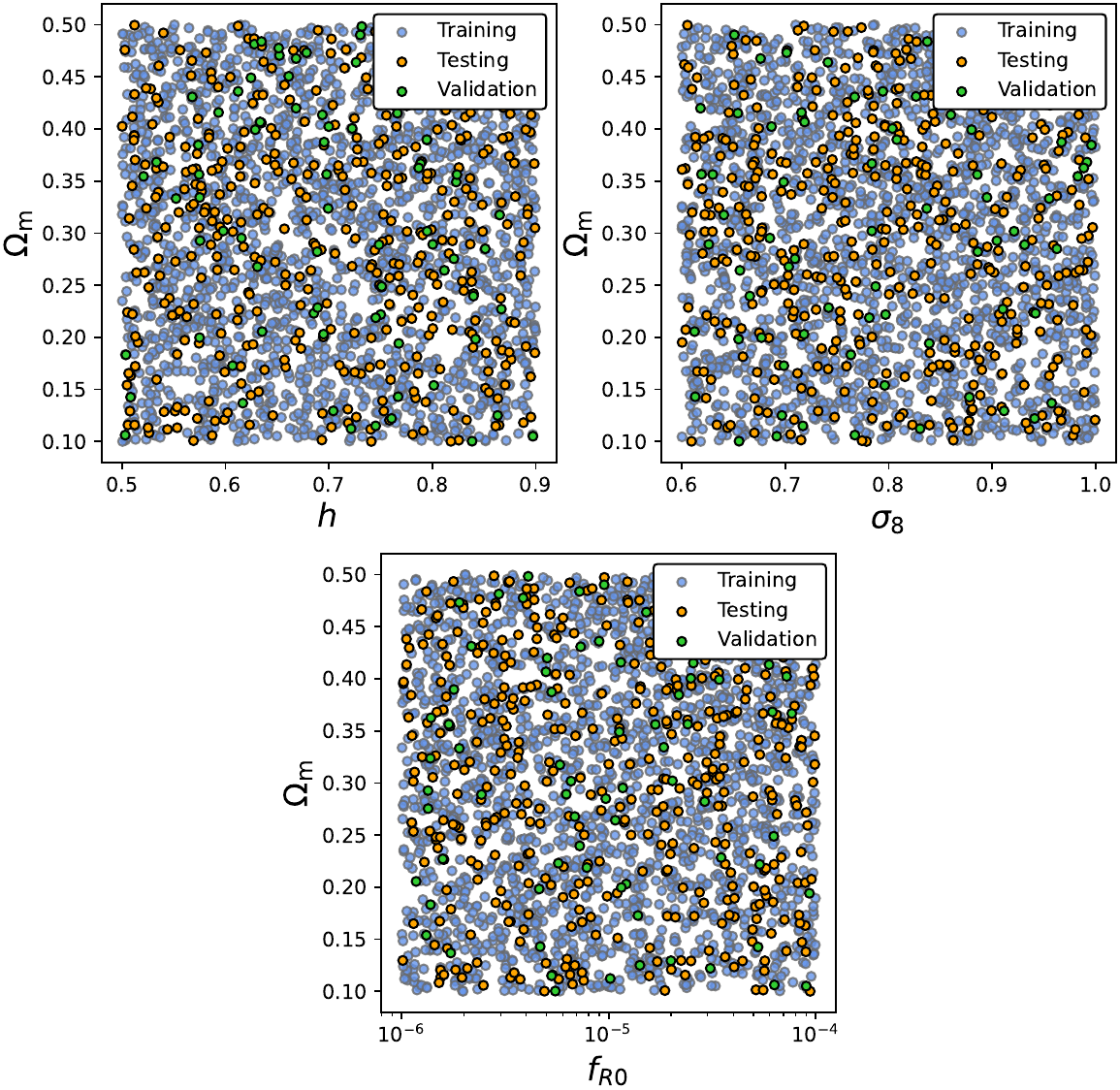}
    \caption{Distribution of the 2500 $f(R)$ cosmologies employed in the diffusion models in selected parameter planes. Light-blue dots represent the training dataset, orange dots the testing dataset, and green dots correspond to the validation dataset. The parameter planes shown are $\Omega_m$ versus $h$, $\sigma_8$, and $f_{R0}$.}
    \label{fig:paramspace}
\end{figure}

We used a dataset already generated by~\citep{2024A&A...684A.100G}. These simulations were created with the COmoving Lagrangian Acceleration (COLA) algorithm~\citep{Tassev_COLA_2013JCAP, Koda_COLA2016}, 
in particular, the authors used \texttt{MG-PICOLA}\footnote{The code can be found at 
\url{https://github.com/HAWinther/MG-PICOLA-PUBLIC}}~\citep{Winther-COLA_scaledependent2017},
a modified version of \texttt{L-PICOLA}~\citep{2015A&C....12..109H} 
that has been extensively tested against full N-body simulations and that extends the gravity solvers to a variety of gravity models
\input{table_cosmosims.tex}. The dataset consists in 2500 modify gravity simulations varying four cosmological parameters $\Theta=\{\Omega_m,\, h,\,\sigma_8,\, f_{R0}\}$, where $h$ is the reduced Hubble parameter, $\sigma_8$ the r.m.s. density fluctuation within a top-hat sphere of $8$ Mpc/$h$ radius and $f_{R0}$ the amplitude of the modified gravity function in the Hu \& Sawicki model (HS)~\citep{HuSawicki_2007}. The remaining cosmological parameters are set to $\Omega_b = 0.048206$ and $n_s = 0.96$, which correspond to the values reported by~\citet{Planck_parameters2020}. 
The parameter space is sampled with random numbers uniformly distributed within the specified ranges for each parameter (see Table~\ref{tab:simsetup}). Fig. \ref{fig:paramspace} shows the distribution of the 2500 $f(R)$ cosmologies used in this work, presented in a plane projection and highlighting the datasets used for training (light blue dots), testing (orange dots) and validation (green dots).
Each simulation follows the dynamics of the particles $128^3$ in a small box of comoving side-length $256$\Mpch,  using 100 timesteps from an initial redshift $z_i=49$ to a redshift $z=0$.  The training set comprises 80\% of the data (and validation), which corresponds to 2000 boxes containing the overdensity fields, while the remaining 20\% of the data was used for testing. For each simulation, we estimate the density field using a cloud-in-cell
(CIC) particle mesh assignment on a grid with $64^3$ voxels. We consider the effective range of the power spectrum up to the Nyquist frequency, $k_\mathrm{Ny}$, which in our simulations corresponds to $k \approx 0.75$ \MpchInv. 

\section{Methodology}
\begin{figure*}
    \centering
\includegraphics[width=\linewidth]{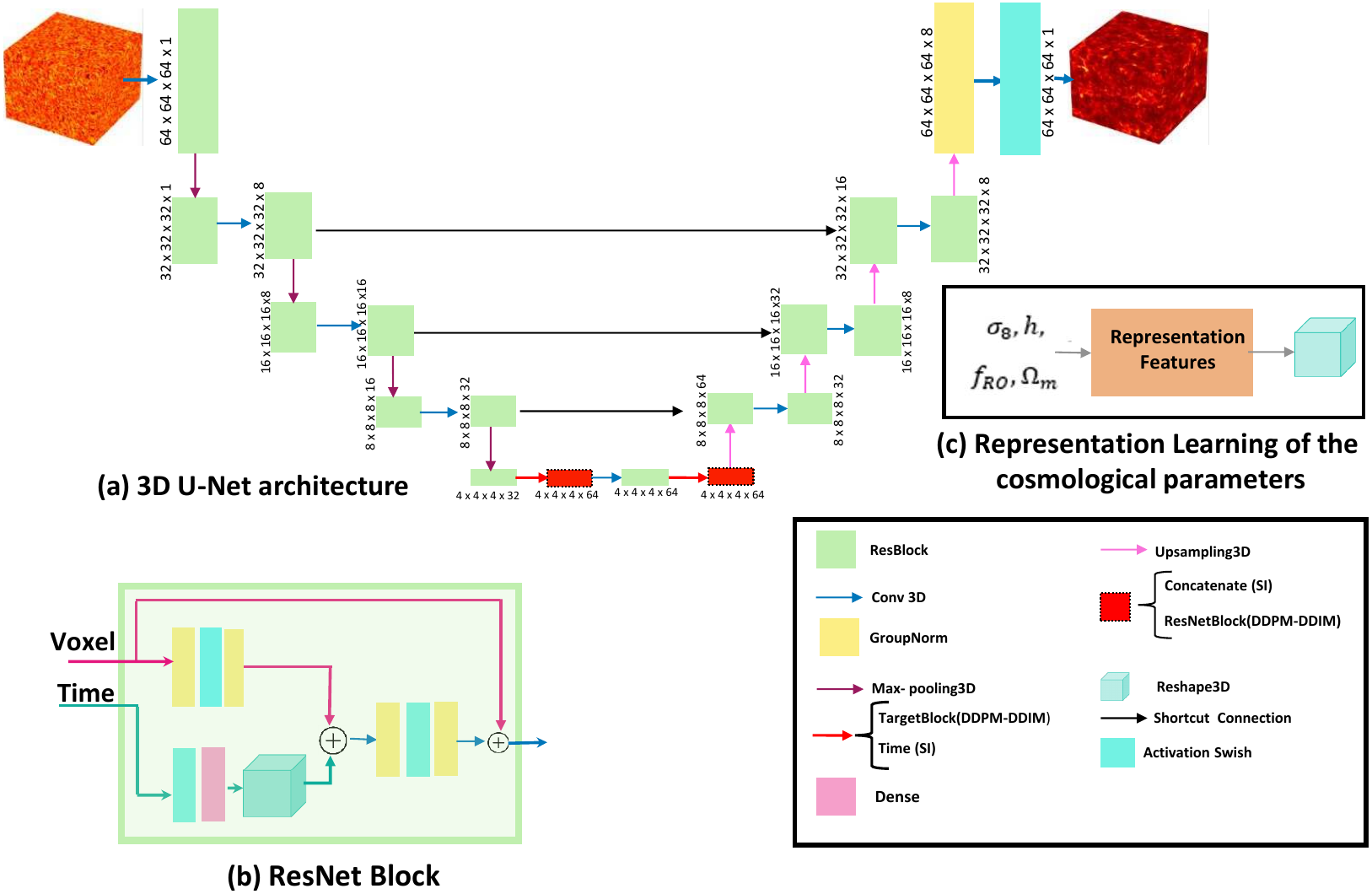}
    \caption{3D UNet modules employed for all models in the paper. (a) UNet architecture composed of ResNet blocks and connections between blocks at each level of the encoder-decoder. (b) ResNet module schema. (c) Target block schema that transform the parameter space into a 3D representation to be inserted into the UNet. }
    \label{unetfig}
\end{figure*}
\begin{figure}
    \centering
    \includegraphics[width=\linewidth]{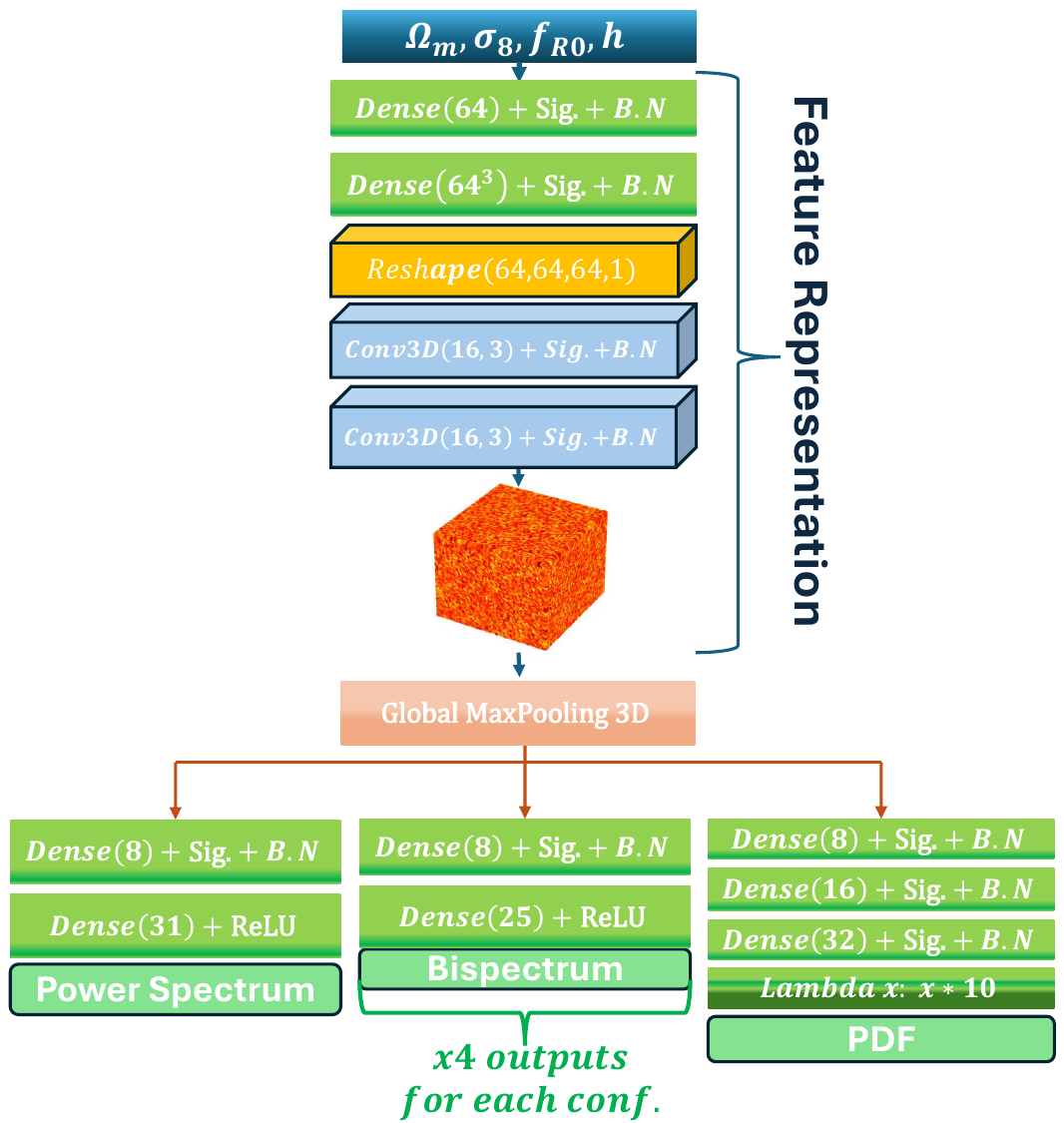}
    \caption{Multioutput regression task for predicting the  power spectra (PS), probability distribution function of the voxels (PDFs), and four different configurations for the bispectra ([Bis$_1$,..., Bis$_4$]) from a set of cosmological parameters. Following the training phase, a submodel is extracted from the layers preceding the Global MaxPool 3D  in the neural network. This submodel enables the generation of a three-dimensional feature representation of the cosmological parameters, capturing their complex relationships and working as a conditioner for the generative models. }
    \label{multifig}
\end{figure}
\subsection{Neural Network Architecture}
While DDPM and DDIM employ neural networks to predict noise at each time step during reverse diffusion, stochastic interpolants use them to estimate the velocity field $\hat{v}(t,x_t)$. The architecture used in this research for all approaches is the 3D-UNet depicted in Fig.~\ref{unetfig}. This model starts with $64\times64\times64$ voxels with 1 channel, which are passed to a calculating schedule across $T=1000$ timesteps, geometrically (cosine) interpolating noise levels from a
Beta Start of $1\times 10^{-4}
$ to a Beta End of $0.02$. Several experiments were performed using lineal, polynomial, and sigmoid functions, however, cosine functions provided the best performance. This UNet consists of an encoder, the middle module, and its decoder. The feature maps of the same pixel level are concatenated via shortcut connections between the encoder and decoder. In the encoder, max-pooling is used to down-sampling layer halves the feature maps, enhancing feature extraction and expanding the receptive field. On the other hand, up-sampling3D in the decoder increases the feature maps, progressively restoring the spatial resolution of the original volume. The output of the decoder part  is processed with a group normalization layer followed by an swish activation and a final convolutional block. In the middle module, we have different configurations depending on the method used (DDPM or SI). For the SI case, four ResNet blocks are sequently used. The ResNet block shown in Fig.~\ref{unetfig}-(b) processes both the feature maps and the timesteps.  The latter is first projected onto an embedding space of dimension 32 using sinusoidal scaling, and then processed through two dense layers of 32 neurons, each with swish activation functions. In case that DDPM (and DDIM) is employed, the middle module consists in two paired ResNet-Target blocks where the cosmological parameter information is inserted into the architecture. Fig.~\ref{unetfig}-(c) illustrates the target block schema, where the cosmological parameters are fed into a pre-trained neural network to get a 3D feature representation (explained later in Subsec.~\ref{feauturerepresentationsection}) of these parameters. The resultant parameter voxel is then concatenated to the feature maps coming from the encoder part.  

\subsection{Feature Representation for Cosmological Parameters}\label{feauturerepresentationsection}
We built a 3D volume feature representation for the cosmological parameters to either aggregate it with the simulation boxes along with the time-steps in DDPM (and DDIM) or define the base density $x0\sim\rho_0$ in the SI approach. We propose to build this parameter volume based on the so-called representation learning, a powerful  technique that enable a neural network to automatically discover and learn the most useful representations of raw data~\citep{bengio2014representationlearningreviewnew}. First, we developed a multioutput regression model using the neural network displayed in Fig.~\ref{multifig}. For this task, we compute the summaries for all train, validation and test volumes such as the power spectra (PS), probability distribution function of the voxels (PDFs), and four different configurations for the bispectra (Bis). Then, we take data pairs ([$\Omega_m$, h, $\sigma_8$, $f_{R0}$],[PS, PDFs, Bis$_1$,..., Bis$_4$]) for training the model in a supervised way. The network receives the cosmological parameters as input, which are then processed by two dense layers, each with 64 neurons, followed by a sigmoid activation function and batch normalization. The output features are reshaped into a $(64,64,64,1)$ volume and passed through three 3D convolutional layers, each employing 16 filters, sigmoid activations, and batch normalization. Then, a 3D convolutional layer with one filter along with a sigmoid activation is applied generating  a 3D representation of the input parameters with dimensions matching the simulation boxes. This sub-neural network yields the 3D representation  used in the diffusion models. Following with the neural network architecture, a 3D global max pooling operation is applied to flatten the volume, resulting in six output branches, each corresponding to one of the pre-defined summaries. The optimized network architecture is shown in Fig.~\ref{multifig}. Training is performed using a Huber loss, with weighted losses assigned to the power spectra and PDF to prioritize their accuracy.

\subsection{Training and summary statistics}\label{secmoments}
Simulation data normalization involved clipping values using the minimum of all maximum values found across the boxes in the training dataset. Subsequently, we applied a logarithmic scale and normalized the data by subtracting the minimum logarithmic value (logmin) and dividing by the range of logarithmic values (logmax - logmin), both calculated from the training data. All models were trained using the Huber loss in Eq.~\ref{equation7} instead of the standard mean squared error. The models were optimized with the Adam optimizer employing a learning rate of $10^{-4}$, a batch size of 16, and training for 30 epochs. Callbacks were implemented to mitigate overfitting. The DDPM model, with approximately 15 million parameters, required approximately nine hours of training on a 16GB Nvidia T4 GPU, while the interpolant model required twelve hours on the same GPU.

\subsubsection{A quality metric for generated density fields: n-point statistics}
The spatial distribution of dark matter  is non-Gaussian, and remarkably little is known about the information encoded in it
about cosmological parameters beyond the power spectrum. Therefore, it is crucial that generative models can learn significant information  well beyond its power spectrum. Therefore, to illustrate the quality generation of our emulators, we compute some summary statistics that provide information about the Gaussian and non-Gaussian signals. We start using the one-point statistics, commonly known as the probability density function (PDF).  The PDF reveals density variations within the simulated volume, identifying overdense regions like galaxy clusters and dark matter halos, as well as underdense regions such as cosmic voids. The values of density contrast $\delta$ are binned using logarithmically spaced bins. The PDF of the cosmic density field is then defined as the normalized number of cells as:
\begin{equation}
P(\delta_i) = \frac{N_i}{N_{\text{total}} \Delta \delta},
\end{equation}
where \(N_i\) is the number of samples in the \(i\)-th bin, \(N_{\text{total}}\) is the total number of samples, and \(\Delta \delta\) is the width of each bin. The next statistical moment is the  matter power spectrum defined as
\begin{equation}
\left\langle\delta\left(\mathbf{k}\right) \delta\left(\mathbf{k}'\right)\right\rangle=(2 \pi)^3 \delta_D\left(\mathbf{k}+\mathbf{k}'\right) P(k),
\end{equation}
where angular brackets denote ensemble average, $\delta_D$ is the 3D Dirac
delta function, which enforces the homogeneity of the density statistics, and $\mathbf{k}$, $\mathbf{k'}$ are Fourier modes. The fact that the power spectrum depends only on the magnitude $k \equiv |k|$ is required by isotropy, which allow us to provide information about the Gaussian signal in the data. In addition to the two-point statistics, we also considered the three-point statistics of the density field. These statistics are able to capture any non-Gaussianities in
the density field. The matter bispectrum $B\left(k_1, k_2,k_3\right)$ is defined as
\begin{equation}
\left\langle\delta\left(\mathbf{k}_1\right) \delta\left(\mathbf{k}_2\right) \delta\left(\mathbf{k}_3\right)\right\rangle=(2 \pi)^3 \delta_D\left(\mathbf{k}_1+\mathbf{k}_2+\mathbf{k}_3\right) B\left(k_1, k_2,k_3\right).
\end{equation}
Unlike the power spectrum, which is only sensitive to the magnitude
of Fourier modes, the bispectrum is the lowest-order correlator that
is sensitive to phases. Because homogeneity constraints the wavenumbers $(k_1 + k_2 + k_3)$ to form a closed triangle, we can also express the bispectrum as a function of two magnitudes and an angle, i.e. $B(k_1, k_2,\theta)$. It is useful, particularly in analyses of modified theories of gravity to consider the reduced bispectrum
\begin{equation}
Q\left(k_1, k_2, k_3\right)=\frac{B\left(k_1, k_2, k_3\right)}{P\left(k_1\right) P\left(k_2\right)+P\left(k_2\right) P\left(k_3\right)+P\left(k_1\right) P\left(k_3\right)},
\end{equation}
to remove the information that is already contained in the power
spectrum. Note that $Q(k_1, k_2,k_3)$ can be written as $Q(k_1, k_2,\theta)$ which define a unique triangle given two out of the three arguments. We use the Pylians3 library to compute these statistics\footnote{\url{https://pylians3.readthedocs.io/}}.

\section{Results}

\begin{table*}\centering
\begin{tabular}{lllllllllll}
\hline
\multicolumn{1}{c}{\textbf{Model}} & \multicolumn{3}{c}{\textbf{Power Spectra}}                                                & \multicolumn{3}{c}{\textbf{\begin{tabular}[c]{@{}c@{}}Bispectra \\ $(k_1=2k_2=0.3)$\end{tabular}}} & \multicolumn{3}{c}{\textbf{PDF}}                                                          & \multicolumn{1}{c}{\textbf{\begin{tabular}[c]{@{}c@{}}Inference\\ Time\end{tabular}}} \\ \cline{2-10}
\rowcolor[HTML]{EFEFEF} 
                                   & \multicolumn{1}{c}{\cellcolor[HTML]{EFEFEF}\textbf{MSE}} & \textbf{MAE}  & \textbf{$R^2$} & \multicolumn{1}{c}{\cellcolor[HTML]{EFEFEF}\textbf{MSE}}    & \textbf{MAE}     & \textbf{$R^2$}    & \multicolumn{1}{c}{\cellcolor[HTML]{EFEFEF}\textbf{MSE}} & \textbf{MAE}  & \textbf{$R^2$} &                                                                                       \\ \hline
\textbf{DDPM}                      & \textbf{61.67}                                           & \textbf{1.70} & \textbf{0.89}  & \textbf{6.95}                                               & \textbf{2.21}    & \textbf{0.81}     & 0.31                                                     & 0.27          & 0.83           & 3m 20.2s.                                                                             \\
\rowcolor[HTML]{EFEFEF} 
\textbf{DDIM}                      & 69.25                                                    & 1.86          & 0.80           & 16.35                                                       & 3.42             & 0.72              & 1.02                                                     & 0.53          & 0.41           & \textbf{9.9s}                                                                         \\
\textbf{SI}                        & 115.14                                                   & 2.29          & 0.72           & 22.56                                                       & 3.97             & 0.64              & \textbf{0.14}                                            & \textbf{0.18} & \textbf{0.89}  & 45.9s                                                                                 \\ \hline
\end{tabular}
\caption{Assessment of the generative models through the test set, with bold values indicating superior performance. The results show that DDPM achieves the lowest error for most statistical moments, outperforming the majority of the models. However, SI performs best in terms of PDF accuracy. Additionally, DDIM stands out by generating synthetic datasets in just 10 seconds. Here  we present the results specifically for the most standard bispectrum configuration $k_1=2k_2$.}\label{tablemetrics}
\end{table*}

\begin{figure*}
    \centering
    \includegraphics[width=\linewidth]{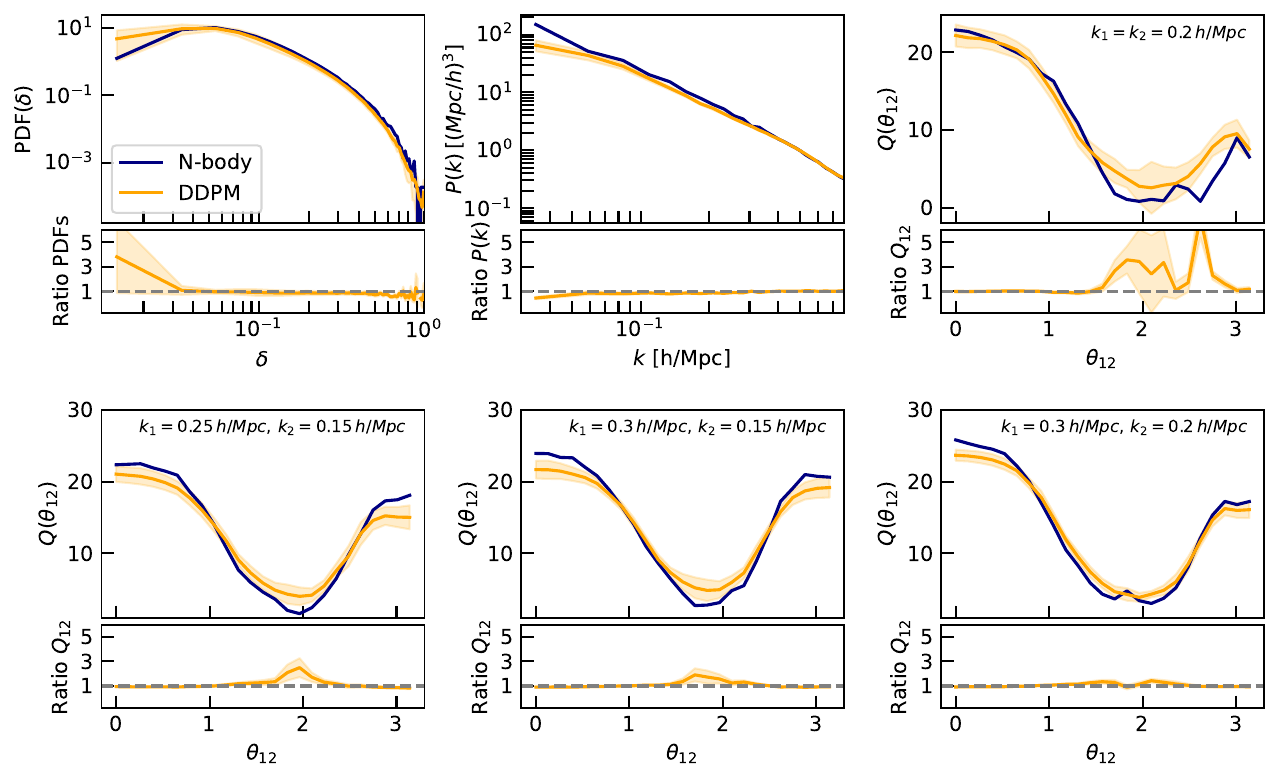}
    \caption{Summary statistics  for generated fields with DDPM corresponding to the fiducial value ($\Omega_m=0.305$,$\sigma_8=0.710$, $h=0.561$, and $0.1\log_{10}|f_{R0}|=0.43$). Each panel shows the probabilistic distribution function PDF with $50$ bins, power spectrum $P(k)$, and four bispectra  configuration $Q(\theta_{12})$. Solid orange line represents the mean over 50 generative samples, while the orange region defines the standard deviation. Bottom plots illustrate the percent error for each summary statistics.  }
    \label{figddpm1}
\end{figure*}
\begin{figure*}
    \centering
    \includegraphics[width=\linewidth]{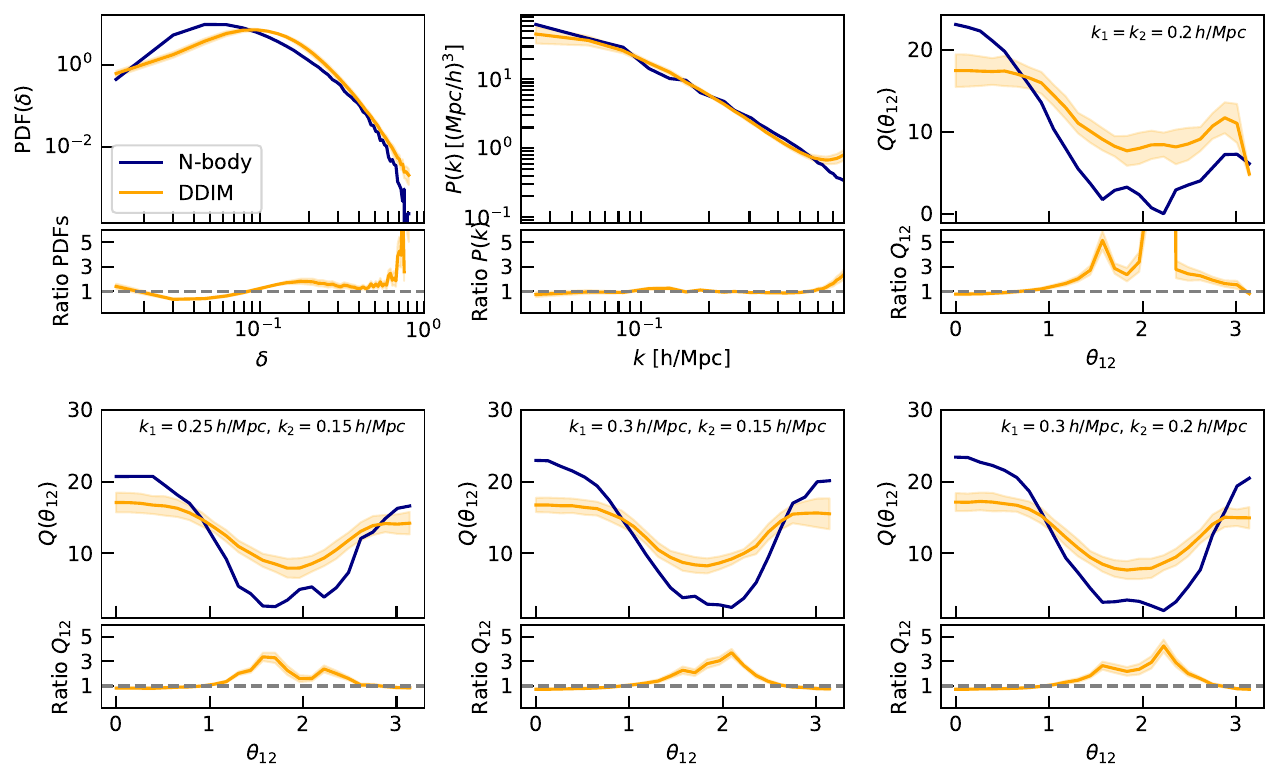}
    \caption{ Summary statistics  for generated fields with DDIM corresponding to the fiducial value ($\Omega=0.5$,$\sigma_8=0.7$, $h=0.2$, and $f_{R0}=0.1$). Each panel shows the probabilistic distribution function PDF with $50$ bins, power spectrum $P(k)$, and four bispectra  configuration $Q(\theta_{12})$. Solid orange line represents the mean over 50 generative samples, while the orange region defines the standard deviation. Bottom plots illustrate the percent error for each summary statistics.}
    \label{figddim1}
\end{figure*}
\begin{figure*}
    \centering
    \includegraphics[width=\linewidth]{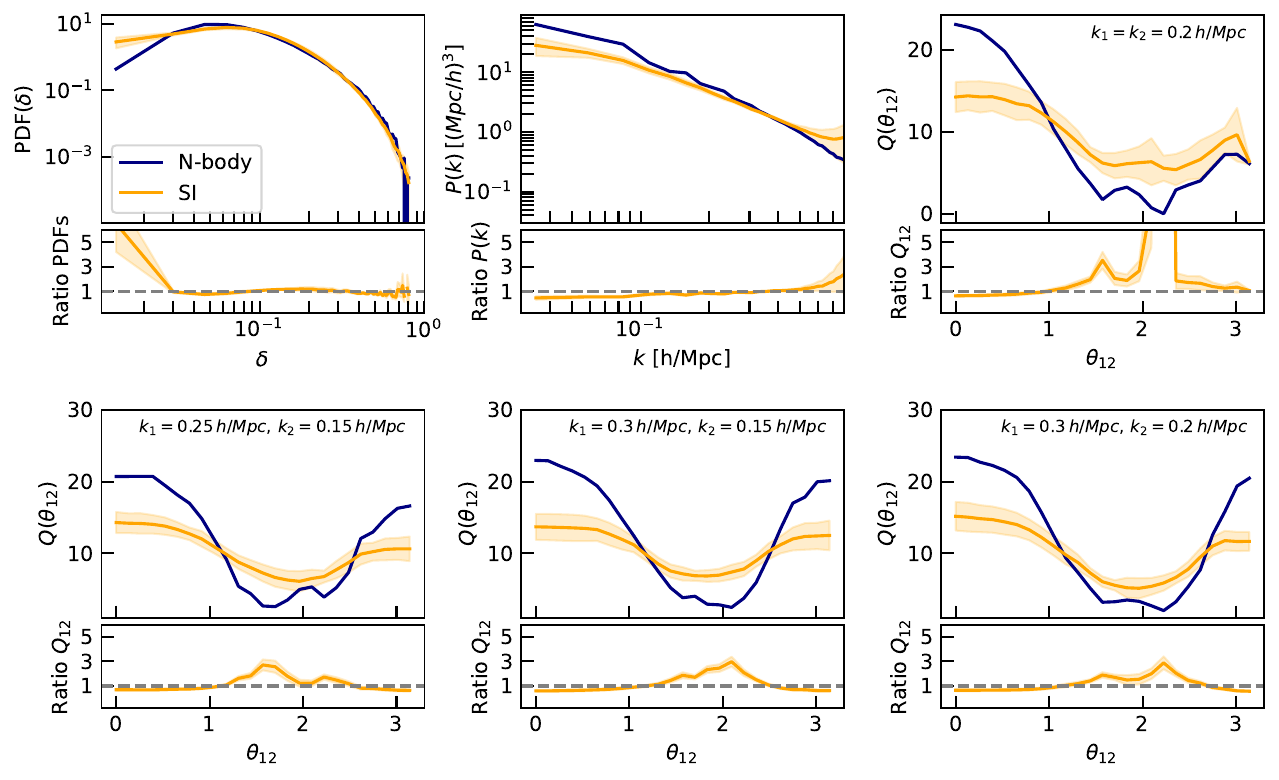}
    \caption{Summary statistics  for generated fields with SI corresponding to the fiducial value ($\Omega=0.5$,$\sigma_8=0.7$, $h=0.2$, and $f_{R0}=0.1$). Each panel shows the probabilistic distribution function PDF with $50$ bins, power spectrum $P(k)$, and four bispectra  configuration $Q(\theta_{12})$. Solid orange line represents the mean over 50 generative samples, while the orange region defines the standard deviation. Bottom plots illustrate the percent error for each summary statistics. }
    \label{figsi1}
\end{figure*}
Having thoroughly examined the methodologies employed for DDPM (and DDIM) as well as SI models, we now turn our attention to evaluating their performance in generating 3D density fields. To assess the efficacy of the DDPM model, we trained it on the relevant dataset and subsequently generated 50 synthetic samples. These samples were then rigorously compared against a test set instance with identical cosmological parameters, focusing on their summary statistics. The outcomes of this analysis for the DDPM model are illustrated in Fig.~\ref{figddpm1}. The results demonstrate that the DDPM-generated fields exhibit remarkable consistency with the true field. Not only do they accurately capture the Gaussian signal, but they also successfully recover a diverse range of bispectra configurations. Specifically, two of the bispectra configurations analysed were directly aligned with those used during the training of the feature representation (as detailed in Sec.~\ref{feauturerepresentationsection}), while the remaining configurations represent interpolations between these key points. This highlights the models ability to generalise and produce physically meaningful outputs, even for configurations not explicitly encountered during training. Notice that for lower wavenumbers($k$),  the predicted power spectrum deviates significantly from the true one. This discrepancy can be attributed to the finite size of the volumes, which inherently imposes a cut-off on large-scale modes. Due to the periodic boundary conditions and the limited spatial extent of the simulation boxes, modes with wavelengths exceeding the box size are effectively truncated. As a result, the power spectrum and bispectra, are influenced by the absence of these large-scale fluctuations. For the latter, we can observe slight deviations in their tails. These limitations are particularly significant in cosmological simulations with a small resolution size, as large-scale modes play a crucial role in shaping the structure of the density fields. As previously discussed, one of the primary drawbacks of DDPM  is the significant computational time required to generate samples. This is due to the iterative nature of the process, where the neural network must denoise the image over $T=1000$ sequential steps. To address this limitation, DDIM was introduced as an alternative approach during the inference process after training the DDPM model. DDIM accelerates the generation process by relaxing the Markovian assumption, as detailed in Sec.~\ref{ddimsec}. While this modification substantially reduces inference time, it comes at the cost of a slight degradation in the quality of the generated samples. This trade-off is evident in Fig.~\ref{figddim1}, where the bispectra of the DDIM-generated simulations show a noticeable, though modest, deterioration compared to those produced by DDPM. The balance between sample quality and inference time is a critical consideration, particularly in applications requiring the bispectra  to constrain cosmological parameters. During the validation phase, this trade-off must be carefully calibrated to ensure that the reduction in computational cost does not compromise the scientific utility of the generated samples. By fine-tuning this balance, DDIM enables the efficient production of a high number of volume samples in a shorter time, making it a practical choice for large-scale simulations despite its minor quality trade-offs. Note that in the DDIM model, the small Fourier modes exhibit behaviour consistent with the ground truth. However, deviations begin to emerge for larger modes, reaching up to 30\% error. This discrepancy can be linked to the slight degradation in quality observed in DDIM, as we typically expect precise reconstruction at scales below the Nyquist frequency. Despite this, the power spectrum can be recovered within tens of percent accuracy across the entire range.
Finally, the statistics of a generated sample from the SI approach are illustrated in Fig.~\ref{figsi1}. It is evident that the SI method yields lower performance compared to the previous model. While SI successfully reconstructs the probability distribution function for the majority of the samples, 
it struggles to accurately capture higher statistical moments, despite displaying favourable trends and shapes relative to the ground truth.In Table~\ref{tablemetrics}, we present the evaluation metrics for all models using the entire test dataset. We employ the mean square error (MSE), mean absolute error (MAE), and the coefficient of determination ($R^2$) to assess the accuracy of the statistical moments derived from the power spectrum, a bispectrum configuration, and the probability density function (PDF). Our results indicate that the Denoising Diffusion Probabilistic Model (DDPM) significantly outperforms the other models, achieving the lowest error across all metrics. However, DDPM requires more time to generate volumes compared to the other methods. In contrast, the Denoising Diffusion Implicit Model (DDIM) generates images in just $9.9s$, making it a practical choice for applications where faster generation is essential such as a parameter constraints. Additionally, we observe that the Stochastic Interpolation (SI) model excels at recovering the PDF and generates data in less than a minute. We think that further refinement through hyperparameter tuning could enhance SI performance, potentially making it a highly accurate and efficient model in terms of both precision and inference time. 
\section{Discussion}
This work has demonstrated the potential of conditional generative modelling to accurately create 3D dark matter density fields, capturing high-order statistical moments with considerable success. Our approach offers a promising avenue for generating realistic cosmological structures, a crucial task for various analyses in cosmology.  The demonstrated consistency with higher-order statistics underscores the model ability to capture the complex, non-Gaussian nature of the cosmic web, a significant improvement over methods that rely solely on two-point statistics. This capability is particularly relevant for studying in future phenomena sensitive to the details of structure formation, such as gravity model, galaxy formation, weak lensing and develop parameter estimation~\citep{2024A&A...690A..27G, ono2024debias,andrianomena2024cosmologicalmultifieldemulator}. However, our current model exhibits limitations, particularly at lower wavenumbers. This increased uncertainty stems from the limited volume of the training simulations.  The relatively small box size restricts the representation of large-scale structures, leading to less accurate predictions on these scales.  This limitation highlights the critical need for training data that encompasses a wider range of scales to capture the full spectrum of cosmic structures.  Future work will therefore prioritize training our models on significantly larger simulations, which will provide access to a broader range of wave modes and improve the model performance in the low-wavenumber regime.  This will be crucial for accurately modelling large-scale structure and its impact on cosmological observables~\citep{sharma2024fieldlevelemulatormodelingbaryonic,Jamieson_2023,2024A&A...690A..27G}.
To address the computational challenges associated with larger simulations, we plan to transition to latent diffusion models in future studies. This approach offers a compelling pathway to enhance both training and generation efficiency.  By learning a latent space that is perceptually equivalent to the simulation space, we can operate in a lower-dimensional space, significantly reducing the computational cost.  The core assumption of latent diffusion, that noise perturbation in simulation and latent spaces are compatible with the generative process, allows for efficient sampling and manipulation of the latent representation.  This will enable us to train on larger and more complex simulations, ultimately leading to a more robust and accurate generative model~\citep{rombach2022highresolutionimagesynthesislatent,podell2023sdxlimprovinglatentdiffusion}. Furthermore, exploring the impact of baryonic physics and different feedback mechanisms is essential for a complete understanding of structure formation.  We intend to extend our analysis by training our models on data from the CAMELS simulation suite~\citep{CAMELS,CMD}, utilizing different astrophysical feedback prescriptions such as those implemented in the IllustrisTNG~\citep{Springel_2017}  simulations.  This will allow us to investigate the influence of baryonic processes on the dark matter distribution and to develop a more comprehensive model of cosmic structure.  Additionally, we plan to incorporate the effects of observational distortions, such as redshift-space distortions and lensing effects, into our model.  This will bring our generated density fields closer to observable quantities and enhance their utility for cosmological analyses. Finally, we are particularly interested in leveraging our generative model for parameter inference.  By combining our model with MCMC techniques, we can potentially constrain cosmological parameters and explore the degeneracy between different cosmological models~\citep{Hort_a_2020,mudur2024diffusionhmcparameterinferencediffusion}.  The ability to generate realistic density fields efficiently opens up new possibilities for exploring the likelihood surface and constrain the cosmological parameters faster.

\section{Conclusions}

In this work, we have explored the application of conditional generative modelling, specifically using Denoising Diffusion Probabilistic Models (DDPMs), along with their accelerated variant DDIMs, and Stochastic Interpolants (SI) to generate 3D dark matter density fields.  Our analysis demonstrates the significant potential of diffusion models for this task.  Our findings demonstrate that DDPM excels in capturing the complex statistical properties of these fields, accurately reproducing both the power spectrum and bispectrum, even for configurations not explicitly encountered during training. However, the computational cost associated with DDPM iterative generation process presents a significant limitation.  While DDIM offers a substantial speed-up in sample generation, it comes at the expense of a slight reduction in accuracy, particularly at larger wave modes.  This trade-off between speed and accuracy is crucial and must be carefully considered depending on the specific application.   Finally, the SI model, while capturing some trends in the bispectrum and successfully reconstructing the probability distribution function for most samples, exhibits lower overall performance compared to both DDPM and DDIM, especially in capturing higher statistical moments.  Our quantitative evaluation confirms the superior performance of DDPM, followed by DDIM, highlighting the potential of diffusion-based generative models for cosmological applications. Future work will focus on mitigating the limitations of DDPMs computational cost, potentially through further refinements of DDIM or exploration of other accelerated sampling techniques, and further exploring the use of these models for parameter inference and other key cosmological tasks.  Additionally, addressing the low-wavenumber discrepancies observed in all models due to finite simulation volume will be a priority, requiring the use of larger training simulations. Our code, and scripts used to
produce the results in this paper can be found at \href{https://github.com/JavierOrjuela/generative-models-f_R_2025}{Github\faGithub}.

\section*{Acknowledgements}
This paper is based on work supported by the Google Cloud Research Credits program with the award GCP19980904 granted to H.J. Hort\'ua. J.E. García-Farieta research was financially supported by the project ``Plan Complementario de I+D+i en el área de Astrofísica'' funded by the European Union within the framework of the Recovery, Transformation and Resilience Plan - NextGenerationEU and by the Regional Government of Andalucía (Reference AST22\_00001).





\bibliographystyle{mnras}
\bibliography{mnras} 





\bsp	
\label{lastpage}
\end{document}

%% file: table_cosmosims.tex
\begin{table}
      \caption[]{The summary of the set-up of the MG simulations. Left: cosmology parameters. Right: set-up parameters used for \texttt{MG-PICOLA} code.}\label{tab:simsetup}
\begin{tabular}{ccccc}
\cline{1-2} \cline{4-5}
\multicolumn{2}{c}{\textbf{Cosmologies}}                                    &  & \multicolumn{2}{c}{\textbf{Simulation setup}}                           \\ \cline{1-2} \cline{4-5} 
\cellcolor[HTML]{EFEFEF}$\Omega_m$ & \cellcolor[HTML]{EFEFEF}{[}0.1, 0.5{]} &  & \cellcolor[HTML]{EFEFEF}Boxsize    & \cellcolor[HTML]{EFEFEF}$256$ \Mpch \\
$h$                                & {[}0.5, 0.9{]}                         &  & $N_p$                              & $128^3$                            \\
\cellcolor[HTML]{EFEFEF}$\sigma_8$ & \cellcolor[HTML]{EFEFEF}{[}0.6, 1.0{]} &  & \cellcolor[HTML]{EFEFEF}Grid force & \cellcolor[HTML]{EFEFEF}$128^3$    \\
$0.1\log_{10}|f_{R0}|$                 & {[}0.4, 0.6{]}                         &  & IC                                 & 2LPT $z_{ini}=49$                  \\
\cellcolor[HTML]{EFEFEF}$\Omega_b$ & \cellcolor[HTML]{EFEFEF}0.0489         &  & \cellcolor[HTML]{EFEFEF}Steps      & \cellcolor[HTML]{EFEFEF}100        \\
$n_s$                              & 0.9665                                 &  & $k_\mathrm{Ny}$                           & 1.58                              \\ 
\cline{1-2} \cline{4-5}
\end{tabular}
\end{table}